\documentclass[authoryear,review,5p]{elsarticle}

\usepackage{amssymb}
\usepackage{url}

\journal{Planetary Space Science}

\begin{document}

\begin{frontmatter}


\title{History of solar wind and space plasma physics revisited}

\author[1]{T.~E.~Girish}
\author[1]{G.~Gopkumar}
\author[2]{P.~E.~Eapen}

\address[1]{Department of Physics, University College,
Thiruvananthapuram 695\,034,
India email:~\url{tegirish5@yahoo.co.in}}

\address[2]{Department of Physics, St. Gregorius College,
Kottarakara 691\,531, Kollam, India}

\begin{abstract}
A paper published by Scottish geophysicist J.A. Broun in 1858
contained several pioneering and remarkable ideas in
solar-terrestrial physics. He could anticipate more or less
correctly the nature and origin of solar wind, solar magnetic
fields, sunspot activity and geomagnetic storms in the middle of
the  19th century. Broun applied the experimental results of the
behaviour of ionised gases in discharge tubes for the first time
to Space Physics which may be considered as the beginning of the
astrophysical plasma physics. In this context he  attempted to
explain the plasma interactions of solar wind with the comet
tails and earth's magnetosphere. Most of the postulates or
hypotheses put forward by Broun in 1858 and later in 1874 was
rediscovered during the 20th century, after the advent of
Space age.
\end{abstract}


\end{frontmatter}

\section{\label{sec:1}Introduction}

Emergence of physical ideas of solar wind and space plasma
physics during the 20th century significantly contributed to our
current knowledge of planetary sciences. The quest for the
causes of aurora and associations of the geomagnetic variations
with solar phenomena like sunspots inspired the development of
solar-terrestrial physics in the 18th and 19th centuries
\citep{Parker1989}. Kristian Birkeland's Terrella experiments
(1896--1913) were considered as the first study of that kind  in
space plasma physics \citep{RypdalBrundtland1997}. Ideas of transient  particle emission from
the Sun were developed since late 19th century
\citep{Goldstein1881,Mitra1952}. However continuous solar wind
flow from the Sun were actively considered by space physicists
only during early 1950's \citep{Biermann1951}. We need to
revisit the present history of solar wind and space plasma
physics when we consider the scientific contributions of John
Allan Broun during the 19th century.

John Allan Broun, the Scottish geophysicist and astronomer
is well known for his discoveries
of semi-annual variations in geomagnetic activity \citep{Broun1848} and
solar sources of recurrent geomagnetic storms before Maunder
\citep{Broun1876}. Broun served as the director of Matherstom magnetic
observatory in Scotland during the years 1842--1848 and
Maharaja's observatory in Trivandrum during the years 1852--1865
\citep{Clerke2004}.
During his service in Trivandrum (situated near magnetic dip
equator) J.A. Broun wrote a remarkable paper in December 1857
which was published in \emph{Philosophical Magazine} as a letter
next year \citep{Broun1858}. This was one hundred years prior to
the publication of the hydrodynamic theory of the solar wind
\citep{Parker1960}.

In this article we will discuss first about the pioneering ideas
of Broun contained in the above paper about the nature and
origin of solar magnetic fields and sunspot activity at a time
when solar physics was just emerging as a research discipline.
We will also explain how Broun perhaps opened the studies on
astrophysical plasma physics when he applied the experimental
results of electricity and magnetism (gas discharge tube
experiments) to space physics. The nature of solar wind type
medium filling the interplanetary space and its interaction with
plasma bodies like comets and earth's magnetosphere as envisaged
by Broun in his paper \citep{Broun1858} during the middle of 19th
century will be then discussed. It is interesting to find that
experimental confirmation to Broun's postulates/hypotheses in
1858 and later in 1874 could be possible only during the second half of 20th
century with the advent of space age.

\section{\label{sec:2}Broun's ideas of solar magnetic fields and
sunspot activity  in the year 1858}

After the discovery of sunspot cycle \citep{Schawabe1844} the
attempts to connect sunspot activity with geomagnetic phenomena
was purely statistical or empirical in nature and lacked
physical models to support the same. In order to explain certain
laws of geomagnetic variations observed in Europe and
Trivandrum, Broun proposed some new ideas about sunspot activity
and solar magnetic fields with some physical insight in his
\emph{Philosophical Magazine} paper \citep{Broun1858}. The
relevant portions from the text of the above paper will be cited
below (given within inverted comas). We have given our
explanatory notes in the modern scientific perspective for each
citation which are numbered serially in the small Roman
numerals. The page numbers refers to the original paper of
Broun.

\begin{enumerate}
\item[(i)] ``page 94: Does not sun act as a magnet, perhaps as
an electromagnet, the currents forming it being within its
atmosphere?''
\begin{itemize}
\item The origin of solar magnetic fields is due to an electromagnetic
induction process involving an electromagnet and currents
flowing in the solar atmosphere. This can be considered as an
early anticipation of dynamo models of astrophysical magnetic
fields by Broun.
\end{itemize}
\item[(ii)] ``page 94: Are not the solar spots disruptions of
the current due to positions of planets in the plane of its
equator?''
\begin{itemize}
\item J.A. Broun was one of the earliest proponents of the theory of
possible physical connections between planetary dynamics,
sunspot activity and origin of solar magnetic fields which was
followed by others in the 19th century \citep{DeLaRueLoewy1865,DeLaRueetal1872}. This is also a
topic of current interest \citep{Hung2007}. Planets in our solar system
are known to revolve the sun in orbits situated close (within
ten degrees) to the heliographic equator \citep{Fairbridge1967}.
\end{itemize}

\item[(iii)] ``page 95: If the sun act as a magnet it is
possible from the analogy of our earth, that its magnetic poles
will not coincide with the poles of rotation; perhaps even the
poles may have unequal forces. In such a case, it might be
expected that the fact could be determined from our magnetic
observations.''

\begin{itemize}
\item Broun's remarkable and intuitive perception of the nature of
large scale solar magnetic fields. During sunspot minimum and
declining phases of the sunspot cycles large scale solar
magnetic fields often resembles a tilted dipole configuration
\citep{Shultz1973,Zirker1977}. North-south asymmetry in the strength of solar polar
magnetic fields and differences in the epochs of reversals in
northern and southern helio hemispheres are well documented in
the solar-terres\-trial literature of recent times \citep{Babcock1959,MakarovMakarova1996}.
Broun's idea of inferring properties of solar magnetic fields
from geomagnetic data is also acceptable in the modern cotext
\citep{Lockwoodetal1999,Loveetal2012}.
\end{itemize}

\end{enumerate}

\section{\label{sec:3} Beginning of astrophysical plasma
physics: Broun's ideas of solar wind and its interaction with
comets and earth's magnetosphere.}

Studies on the electric discharge lamps and electric arcs
\citep{Crowther1969,Maecker2009} inspired
the investigations on the discharge of electricity through
rarefied gases in the 19th century \citep{Thomson2005}. This
later contributed to the development of atomic physics and
plasma physics. J.A. Broun was perhaps the first physicist to
apply the results of discharge tube experiments in
electromagnetism to space physics which may be considered as the
beginning of astrophysical plasma physics. In this section we
will describe about these pioneering ideas of Broun published in
his philosophical magazine paper\break \citep{Broun1858}. As in the
previous section citations of relevant passages on the above
paper will be presented here with our brief explanatory notes.

\begin{enumerate}

\item[(iv)] ``page 96: Sir John Herschel, I believe has
somewhere suggested electricity as the cause which directs the
tail of comets\ldots. Are not comets formed by magnetic gases? Is
not the tail of the comet due to the directive action of the
solar magnet, the curvature of the tail, sometimes seen, being
due to the position of solar magnetic poles relatively to the
path of the comet? Is not the condensation of comet, when
approaching the sun, a phenomenon similar to those observed by
Dr. Faraday and M. Pl\"ucker in their recent researches on the
action of the poles of a magnet on certain gases or liquids?.''

\begin{itemize}
\item Scientists prior to Broun suggested a polar force probably of
electrical nature from the sun acting on the comet tails during
its passage towards the sun \citep{Mendis2006}. Broun proposes that the
observed changes in the tail of comets may be due to the action
of extended solar magnetic fields in the interplanetary space.
He considers this phenomena analogous to the behavi\-our of
ionised gases like hydrogen in electric discharge tubes under the
action of external magnetic fields as observed by scientists of
his time like Pl\"ucker in Germany and Micheal Faraday in England
\citep{Plucker1858}.

This is an interesting example of the Broun's intuitive guess of
the interaction of interplanetary magnetic field (carried by the
solar wind) with the cometary plasma at least hundred years
prior to the satellite era during which it was experimentally
confirmed \citep{Biermann1963}.
\end{itemize}

\item[(v)] ``page 97: Is not the zodiacal light the magnetic
aether in a luminous state, repelled by the solar magnetic
poles?

Does not zodiacal light revolve around the sun?

If so, what is its period of revolution?

Are not the extent and the intensity of zodiacal light related
to the periods of spots as Cassini and Mairan supposed?''

\begin{itemize}
\item Scientists of the 19th century invoked the idea of an \emph{ether}
type media filling the interplanetary space to felicitate any
kind of energy (example: sunlight) or physical action propagating
through the same. J.A.~Broun also shared a similar idea to
explain the physical action of solar magnetic fields on distant
objects in interplanetary medium such as cometary plasma. If one
assumes an earth like magnetic field existing on the sun with a
strength around $0.4\times 10^{-4}~{\rm T}$ it will be reduced
to $2.5\times 10^{-17}~{\rm T}$ near earth by inverse
square law variations. Like \citet{Mairan1754} Broun also believed
that zodiacal light is a kind of gaseous discharge from the sun.
According to him it is a magnetic material filling the
interplanetary space and is released from the sun due to
repulsive action of the solar magnetic fields. Broun also
assumes that such a gaseous discharge co-rotates with sun
implying that is possibly a continuous outflow from our star. He
also suggests that the density of the above interplanetary
material (as known from intensity and size variations of the
zodiacal light).
can possibly change with the phase of the sunspot activity cycle.
\end{itemize}

\item[(vii)] ``page 97: From the known action of the sun on gases of
comets, may we not infer some action of the sun on the gases
forming our own atmosphere?\ldots\ 
Should not the sun acting as a magnet on the magnetic gases forming
our atmosphere, and by induction on the terrestrial magnet, cause
the atmosphere to assume an ellipsoidal form, having the greater
axis in or near the plane of equator?\ldots\ 
If the form which the atmosphere assumed, under the
influence of terrestrial and solar magnets, were some what
irregular (as in some figures assumed by magnetic liquids,
between poles of a magnet as in M.~Pl\"ucker's experiments)\ldots

\begin{itemize}
\item Eventhough several models of the possible interactions of solar
wind particles with earth's magnetic field were developed during
the first half of the 20th century \citep[see][for details]{Mitra1952}
 the deformation of the earth's magnetosphere from the
dipole geometry (day side compression and extended magnetic
tail) due to interaction with the continuous solar wind flow was
not known until the spacecraft age \citep{KivelsonRussel1995}.
But Broun could anticipate this discovery at least hundred years
prior to its experimental confirmation. He could also infer that
the physical mechanisms behind interaction of extended solar
magnetic fields (carried by the solar wind) with comets and
earth's magnetic atmosphere are identical or similar in nature.

It is well known that search for the physical nature of cathode
rays lead to the discovery of electrons by J.J. Thompson in 1897
\citep{Thompson1897}. Actually cathode rays was discovered much earlier
through pioneering experiments carried by \citep{Plucker1858} on the
action of external magnetic field on rarefied gases in discharge
tubes. The shape of light glows in these discharge tubes
(Geisslers tubes) used by Pl\"ucker for the above experiments surprisingly
resembled the shape of earth's magnetosphere (see Fig.~\ref{fig:1}).
Pl\"ucker's experimental results served as an inspiration to J.A.
Broun to model the interaction of extended solar magnetic fields
(propagating through a gaseous \emph{ether} media in interplanetary
space as assumed by him) with the earth's magnetic atmosphere
(or magnetosphere). According to Broun the shape of the earth's
magnetosphere is deformed and resembles an ellipsoid with
greater axis near the geomagnetic equator.
\end{itemize}

\begin{figure*}
\includegraphics{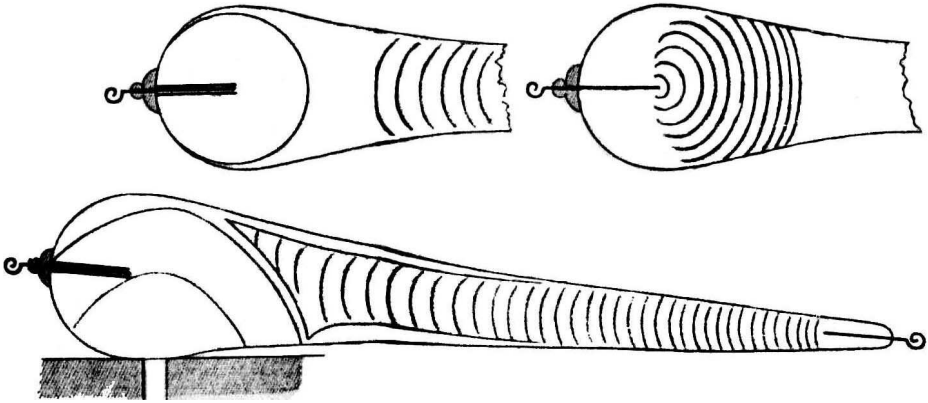}
\caption{The Geissler tubes and glows shown as fringe type
patterns inside the same used by Pl\"ucker in his discharge tube
experiments (part of plate I of Pl\"ucker, 1858)}
\label{fig:1}
\end{figure*}

\item[(viii)] ``page 95: Is not the case that the magnetic
disturbances coexists always with the spots; but it is not
impossible that during the \emph{formation} of the spots the
disturbance is produced that is to say at the period when the
supposed \emph{discharge} of the sun's electrical atmosphere occurs
(words in italics form part of original text).

\begin{itemize}
\item Broun makes same comments on the solar origin of geomagnetic
storms. He suggests two possibilities to explain the observed
association between sunspot activity and magnetic storms. It can
be related to the phenomenon of formation of sunspots itself.
Another possibility is that magnetic storms in earth occurs when
an electrical discharge from the solar atmosphere occurs. The
later suggestion of Broun is similar to the model of
\citet{Goldstein1881} who assumed particle emission from sunspots as the cause
of the observation of aurora on earth.
\end{itemize}

\item[(ix)] ``page 94: After a magnetic disturbance there is a
diminition of force shown on earth, which remains for some days
as if there had been a violent action with the result of loss of
energy.''

\begin{itemize}
\item Geomagnetic storm occurrences in the mid/low latitudes is
characterised by a significant decrease in the horizontal
component of the earth's magnetic field lasting upto 2--3 days
during the main phase of such magnetic disturbances. They are now
understood to be part of large energy dissipation events in the
magnetosphere--iono\-sphere systems as a consequence of solar
wind-magnetosphere energy coupling \citep{Akasofu1981}. It is
quite interesting to find that Broun could anticipate long ago
(in 1858) that geomagnetic storms are indeed part of a huge energy
dissipation phenomena in the earth environment involving strong
or violent physical interactions.
\end{itemize}
\end{enumerate}

\section{\label{sec:4}Discussion}

Several decades prior to the dawn of modern atomic physics and
plasma physics, Broun could propose several pioneering ideas in
solar-terrestrial physics which were re-discovered during the
20th century (Table~\ref{tab:1}).
J.A. Broun was a versatile genius who could think much ahead of his
time. As a visionary he could find striking similarities between
interaction of solar magnetic fields with cometary or
magnetospheric plasma and ionised gas behaviour under external
magnetic fields in discharge tube experiments. This perhaps
opened a new branch of study `astrophysical plasma physics'
surprisingly during the middle of 19th century.

\begin{table*}
\caption{Pioneering ideas of J.A. Broun in 1858 and its rediscovery during the 20th century}
\label{tab:1}
\begin{tabular}{lp{3in}p{3in}}
\hline
&Broun's ideas & Rediscovery during the 20th century\\
\hline
1. &Nature and origin of solar wind
\citep[with additions in]{Broun1874}
& Hydrodynamic theory \citep{Parker1960}
Satellite observations   during  1958--62\\
2. & Sunspot activity:
possible connection with planetary dynamics in the solar system &
\citet{Shuster1905}\\
3. & Nature of large scale solar magnetic fields:-\\
  & (a) origin due to an electromagnetic induction process in the sun
   & Dynamo theory of \citet{Larmor1919}\\
   & (b) tilted dipole configuration & \citep{Shultz1973}\\
	 & (c) north-south asymmetry in solar polar magnetic fields
   & \citep{Babcock1959}\\
4. & Solar wind and IMF  plasma interactions:\\
   & (a) with earth's magnetosphere causing distortion in shape
     & Explorer satellite observations \par \citep{KivelsonRussel1995}\\
   & (b) with comet tails & \citet{Biermann1963},
   \citet{NeidnerBrandt1978}, \citet{Neidner1982}\\
5. & Geomagnetic storm as a  significant  energy  dissipation phenomena in the terrestrial environment
   & Solar-wind magnetosphere energy coupling models \citet{Akasofu1981}\\
\hline
\end{tabular}
\end{table*}

Scientists of the 18th century like \citet{Mairan1754} and early
19th century like Cassini suggested that interplanetary space is
filled with matter of solar origin and Zodiacal light (now
understood to be an optical phenomena in earth's atmosphere) is
visible manifestation of the same. \citet{Broun1858} provided a
physical basis to these ideas by proposing that `magnetic gases'
are continuously flowing outward from the solar atmosphere to
the interplanetary medium by the repulsive action of solar
magnetic fields on these gases\ldots Later in 1874 during the
discussion of Trivandrum geomagnetic observations he proposed a
thermodynamical mechanism of coronal expansion to account for
the solar wind outflow from the sun which is cited below \citep{Broun1874}.


``I believe the facts now certainly connected with solar and
lunar magnetic action cannot be explained without the existence
of an electric medium similar to through it may be denser than
in which light and heat are propagated.

\emph{As a footnote}: The idea of such a medium was suggested by
Herchel in 1843 \citep[see for eg.][]{Steven2004}. It is quite
possible that there may be denser medium around the sun. I have
long ago suggested that the zodiacal light is formed in this
medium by the passage of electric fluid \emph{\citep{Broun1858}}.
If M. Fayes theory of the repulsive action of solar heat, which
blows out the cometary gases so many millions of miles could be
assumed the same action should blow off a part of solar gases.''

\section{\label{sec:5}Conclusions}

\begin{enumerate}[1.]
\item John Allan Broun, the Scottish geophysicist could propose
several pioneering hypotheses/postulates in solar terrestrial
physics which are published in 1858 and 1874.

\item Broun could  suggest   physical mechanisms  for
continuous solar wind flow and solar  coronal expansion in the
above years.

\item Broun was responsible for the beginning of Space
Plasma Physics in the  middle of 19th century when we applied
the  experimental results of the behavior ionized gases in
discharge tubes to Space physics. He could explain solar wind
interactions with come tails and earth's magnetosphere much
before the Space age and dawn of plasma physics. This preceded
Birkeland's Terrella experiments in laboratory space physics by at
least fourty years.

\item Broun's postulates about nature and  origin of solar
magnetic fields, sunspot activity and geomagnetic storms in 1858
were in agreement with modern findings during the 20th century.

\end{enumerate}

\end{document}